# Feasibility Study for CubeSat Based Trusted Node Configuration Global QKD Network

Ekin Aykut*, Hashir Kuniyil, Melis Pahalı, Gizem Gül, Naser Jam, Kadir Durak

Quantum Optics Laboratory, Electrical and Electronics Engineering Department,
Özyeğin University,
Istanbul, Turkey
* ekin.aykut@ozyegin.edu.tr

*Abstract*—Quantum key distribution (QKD) is the most used protocol in the context of quantum cryptography for sharing a private encryption key between two parties. Covid-19 pandemic has raised the ever-increasing need for online communications a lot; this requires enhanced security protocols. QKD has the potential to meet a global scale network's security requirements. Despite considerable progress, all ground-based QKD approaches have distance limitations due to atmospheric or fiber attenuation. A global network scheme can use intersatellite links to establish a trusted node network with constellations. This enables key elements for quantum internet which allows secure exchange of information between quantum computers. The most cost-effective and iterative approach for this goal is to exploit CubeSats. This paper summarizes technical challenges and possible solutions to enable a global QKD network using CubeSats. We discuss practical concerns and alternative paths involved with implementing such systems.

*Keywords—QKD, entanglement, CubeSat, SPDC*

## I. INTRODUCTION

Increasing demand for security-enabled communication services on the ground and in the sky necessitates adoption of new tools and systems and the Covid-19 pandemic raised this as more and more operations are performed online or remotely by humans, robots, and other machines. Among the challenges associated with realization of security-enabled communication services, sharing a secret key is the most critical one and requires access to a number of trusted parties. Quantum key distribution (QKD) can do this with high level of security using axioms from quantum mechanics. Large scale QKD networks need direct line-of-sight optical links that are hard to realize on the ground due to excess loss and noise in long distances typically needed at the global scale. Elevated altitudes can be a solution to this problem and CubeSat is a technology that promises relatively cheap and scalable solution to challenges of QKD [1].

This paper considers challenges in deploying a global QKD network employing CubeSats and addresses some of them. The paper is structured as follows. In Section II, we will briefly review constraints and challenges of intersatellite QKD. In Section III, protocol related concerns will be addressed. Finally, Section IV concludes the paper and gives ideas for further investigation.

## II. CONSTRAINTS AND CHALLENGES OF INTERSATELLITE QKD

CubeSats are proven to be ideal hosts of quantum payloads for establishing large scale QKD networks, due to their cost-effectiveness and ease of design [2]. However, the trade-off of using such a technology is to miniaturize the payload to fit into CubeSat standards, i.e. the size, weight and power (SWaP) requirements of a CubeSat should be considered. The SWaP requirements also put constraints on the optical and RF link budgets, computational power of the embedded system.

### A. Optical Link

Correct modelling and subsequent optimization of the optical link is critical for any optical communication system. Appropriate estimation of loss factors effecting the link quality allows for improvement of the link. The link quality determines the communication bandwidth and error rate directly. In a laser based classical free space optical communication, link quality is dependent on transmitter/receiver efficiencies, pointing accuracy and distance. In such systems it is possible to diverge the beam to compensate for the pointing accuracy related losses. However, since each photon carries information specific to itself in QKD, perfect mode overlap between transmitter and receiver is necessary.

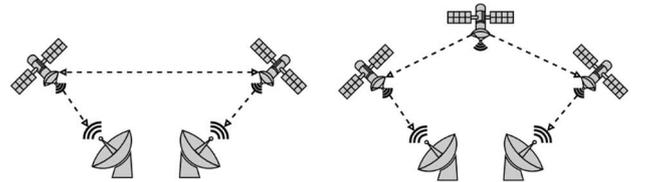

Fig 1. Left: Simple intersatellite QKD network topology. Right: Layered intersatellite QKD network topology.

Transmitter and receiver efficiencies are statistical average values for a SPDC and photon counting based QKD protocol. SPDC process generates photon pairs probabilistically, so effective photon rate can change with an average value during operation. Similarly, photon detection by APDs is a probabilistic process.

Laser beams are generally considered as Gaussian beams with intensity distribution on a surface perpendicular to propagation. With distance, Gaussian beams diverge with an angle $\theta$ depending on beam waist $\omega_0$, thus the aperture size of the optical telescope assembly, and wavelength $\lambda$ as $\theta = \frac{\lambda}{\pi \omega_0}$ and beam width is $\omega(z) = \omega_0 \sqrt{1 + \left(\frac{z}{\omega_0/\theta}\right)}$ .

A highly collimated beam is desired as the beam gets wider, the photon distribution also gets wider and more photons fall outside of receiver telescope at a distance. However, the drawback of a narrower beam is that the received intensity loss is more dependent on the pointing uncertainty due to the centric nature of Gaussian distribution which can be written as below with peak intensity:

$$I(r,z) = I_0 \left( \frac{\omega_0}{\omega(z)} \exp\left(-\frac{2r^2}{\omega(z)^2}\right) \right). \quad (1)$$

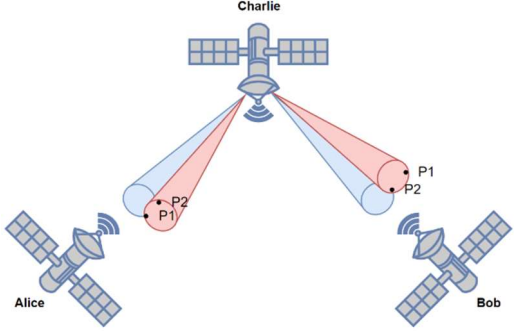

Fig. 2. Red beam demonstrates the mismatch in detected photons with a pointing error.

In order to model the effects of distance and pointing error losses on key rate, we need to consider coincidence probabilities. If we employ a more complex intersatellite QKD scheme such as in Fig. 1, with a satellite distributing entangled photon pairs for QKD between two satellites instead of the generic scheme where the two satellites share a quantum key directly, we encounter another concern regarding the actual coincidences. According to preservation of momentum in SPDC process, the photons of a pair in their respective beams are symmetrical with respect to the center as in Fig. 2. This results in a dramatic decrease and an eventual loss of coinciding pairs with pointing errors as one of the photons fall outside of the receiving telescope. This effect can be examined in the Fig. 3 below. The solution to this problem is simple: If one of the beams is inverted through a confocal lens structure, the beams become equivalent in photon distributions.

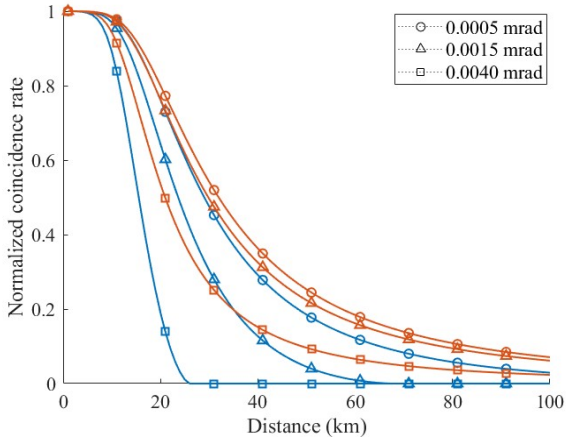

Fig. 3 Blue lines represent the coincidence probability with the effect mentioned in the above paragraph. Red lines represent the coincidence probability with that effect compensated.

The photon source for establishing a QKD network is realized by the spontaneous parametric down-conversion (SPDC) process. SPDC utilizes second-order nonlinear susceptibility tensor $\chi^{(2)}$ in a suitable material. Mostly, satellite-based QKD utilizes $\beta$-barium borate (BBO) crystals as the nonlinear medium because of its unaltered performance against the change of temperature. The other commonly used nonlinear mediums are periodically poled KTP and LN crystals, which require temperature tuning to obtain the desirable entanglement source characteristics; this is undesirable in satellite based QKD. The photon pairs produced in BBO crystal have high-frequency bandwidth (low coherence time) compared to periodically poled crystals, effectively reducing the magnitude of the normalized second-order correlation function $g^{(2)}$. Low $g^{(2)}$ increases noise collection in the system. Figure 1 shows the change in the magnitude of the second-order correlation at zero delay with different source bandwidth values. The problem originated as the $g^{(2)}$ spectrum obtained at the time stamp involves the convolution of the correlation function and the detector impulse response (detector function) which can be characterized by

$$g^{(2)}(\tau) = e^{\left(\frac{\tau_w}{\tau_c}\right)}[f_+(\tau) + f_-(\tau)] \quad (2)$$

where R is the photon pair rate, $\tau_c$ is the coherence time of the source that converts bandwidth $\Delta v$ by $\tau_c = 1/2\pi\Delta v$, $\tau_w$ is the combined effect of detector timing jitter $\tau_j$ and timestamp coincidence window $\tau_b$ that defines as $\tau_w = \sqrt{2\tau_j^2 + (\tau_b/2)^2}$ and $[f_+(\tau) + f_-(\tau)]$ is the error function characterized by $f_\pm(\tau) = e^{\pm\tau/\tau_c}\left[1 \mp erf\left(\frac{\tau \pm \tau_w^2/\tau_c}{\sqrt{2}\tau_w}\right)\right]$ and $erf(x)$ is the error function. The timing jitter effectively creates uncertainty in the correlation time along with the reduced magnitude of second-order correlation as a result of the high bandwidth of the photon pair source created in BBO crystal in the parametric down-conversion process. The large bandwidth of SPDC photons limits the maximum applicable spectral filtering for daylight operation. In this context, non-critical phase matching crystals are preferrable due to their narrow bandwidth photons. However, the challenge of temperature control of crystal makes critically phase-matched crystals still better candidates, like BBO.

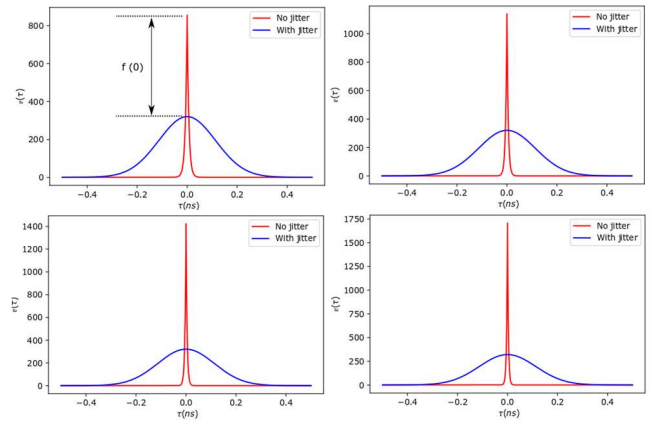

Fig. 4 Plots were obtained by considering the pair source having bandwidths 300, 400 MHz, 500 MHz and 600 GHz. The difference in the magnitude f (0) of the $g^{(2)}$ is increasing with bandwidth.

### A. Attitude Control and Pointing Accuracy Requirements

A CubeSat-based QKD system has fundamentally the same limitations of an optical satellite communication link, i.e., the pointing accuracy. In a typical CubeSat optical transmission system, the main sources of signal loss are geometrical loss and pointing loss. For the geometric part, the link gain which is calculated according to the antenna theory is directly proportional to aperture size of the CubeSat and the ground station telescopes. Size of a typical CubeSat's telescope is limited by the bus size (10 cm) which limits the diameter of a typical onboard telescopes to be around 8 cm.

The size of the ground station's telescope is constrained by atmospheric turbulence. The main portion of geometrical loss is attributed to the CubeSat's telescope size. In the design consideration of an intersatellite link, the two main sources of losses in the link are:

1. Precise orbit determination (POD) and angle accuracy in pitch and yaw.
2. Accurate beam pointing.

For the POD, despite high doppler shift of L-band RF signals due to high speed of CubeSats, our solution is based on combination of Global Navigation Satellite Systems (GNSS) including the US GPS, the Russian GLONASS, the European Galileo and the Chinese BeiDou. Previous works show that with some modifications in the firmware of low-cost, low-power available commercial off-the-shelf (COTS) devices, it is possible to reach fine positioning precision of 0.1 m and velocity precisions of 0.1 m/s [5, 6]. The stabilized non-rotating platform of the CubeSat in this design is an advantage for the GNSS subsystem which provides continuous operation without loss of synchronization without the need to re-acquisition [4]. For CubeSats, two degrees of freedom accuracy in pitch and yaw there is no need for modification and current technologies can be used that generally provide around 50 microradians of pointing uncertainty.

The next challenge in this design is accurate beam pointing. The pointing inaccuracy is the primary cause of loss in the link. It increases bit error rate of single photons. To alleviate the problem of inaccurate beam pointing, in addition to using fast steering mirror (FSM) after telescope, it is possible to use an innovative improvement that is based on high density matrix of Geiger-mode-operated avalanche photodiodes also known as Silicon photomultipliers [7].

Silicon photomultiplier (SiPM) is a solid-state photodetector made of an array of hundreds or thousands of integrated single-photon avalanche diodes (SPADs), called microcells or pixels. By using SiPM instead of single APD in the receiver side, a considerable improvement in pointing requirements can be achieved because this way there is no need to tune the FSM with such a precision to assure a single photon lands on the active area of a single APD. In this design instead, there are few thousands photodetectors with milimeter size effective photosensitive area that increases the photon capture probability by orders of magnitude even in extreme and poor pointing conditions. Such an improvement can be achieved with a product like S13720 from Hamamatsu which has good sensitivity around 760 nm.

### III. Protocol Related Concerns

QKD protocols can be classified into two main categories according to the quantum property they exploit:

1- Prepare and measure (PaM) protocols: In quantum physics, any measurement made on a quantum state causes it to "collapse", i.e. changes the initial state itself. Qubits behave in this manner: each measurement made on Alice's sequence alters it. Eavesdroppers trying to intercept Alice's message would inevitably leave a trace on the sequence, thus the error rate of Bob's measurements would become unusually large, reveling the presence of third parties.

2- Entanglement based: Entangled particles are born in a way such that their quantum states are described in a combined state, i.e. the state of each particle depends on the other, regardless of the distance in between, making this type of protocols are very favorable in larger networks, especially in satellite systems. In such protocols Alice and Bob use entangled pairs distributed by an entangled particle generator. Measuring any entangled particle would change the state of the overall system, therefore if any eavesdropper tries to intercept and measure the sequence of Alice, it alters the system, revealing the interception attempt.

Although PaM protocols are more commonly used, entanglement based networks have the advantage as it allows other quantum protocols that can be used to interact distant stationary qubits.

#### A. Data Link Budget and Key Rate

In order to extract the actual key from the received photons, several post processing steps must be taken. First, a sifted key is obtained by determining the coincidences using the time correlation between photon pairs produced by the SPDC process. After determining the coincidences, measurement basis reconciliation made on the public classical channel and only the photons that are measured in the same basis are used as the sifted key.

Due to strong correlation between the photon pairs, cross-correlation function of the detected photons will have a peak at the time delay value between two receiving nodes. Photons with time difference equal to that delay are considered as coincidences. Cross-correlation between the two functions is:

$$[f(t) \star g(t)](\tau) = \overline{[f(-t)} \ast g(t)](\tau) \quad (3)$$
$$F\{\overline{f(-t)}\} = \overline{F\{f(t)\}} \quad (4)$$

Utilizing the property in Eq. (2), FFT of the signals can be used to compute the cross-correlation in a more efficient way. In general time complexity of the cross-correlation function in time domain is $O(N^2)$ and in the Fourier domain it is $O(N \cdot log(N))$. For long signals and low computational power, efficiency in computing the cross-correlation is vital for the performance of the QKD protocol. Cross-correlation computation of two signals of 100 μs and 500 photons in this window, with 1.296 ns time bins takes around 30 minutes with a program written in C. Same cross-correlation is computed in 0.021 seconds with SciPy library using an FFT based method.

There are two parameters to be optimized while computing the cross-correlation functions, time bin size and coincidence window. Time bin size determines the quantization of the timestamps and the fundamental time unit of the QKD protocol. This quantization has three effects. First, it determines the maximum bit rate that can be achieved as it limits the photon count rate. Only one photon is counted in each time bin. Second, it affects the security of the QKD protocol from a timing jitter side channel attack, which will be explained in the next section. Third, it reduces the required classical communication bandwidth since the quantized timestamps have less bits. In intersatellite communications, data link budget can be as important and can be directly related to power budget.

Coincidence window is the time period in which two measured photons are counted as coincidences after time synchronization between two parties. This window is needed due to timing jitter of the photon detectors. As can be seen in Figure 5, a larger coincidence window results in detection of more coincidences. However, it also increases the rate of accidental coincidences. Accidental coincidences may introduce bit errors in the shared key and reduce S-value of CHSH inequality violation check.

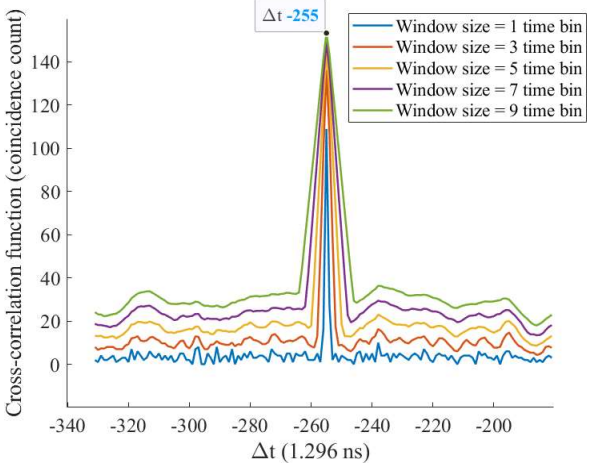

Fig. 5  Cross-correlation of two detector signals with peak at time delay (4096 ns) between the signals

Any bit errors in the key due to non-ideal entanglement fidelity, detected double pairs due to probabilistic nature of the SPDC procedure, or other factors must be corrected using the public channel without compromising the security of the key. There are several methods for error correction proposed in the literature [8]. The most common one being the parity check based cascade, other methods include Hamming code-based Winnow protocol, and LDPC codes commonly used for error correction in classical high-speed communications.

During the error correction process, some information about the key is leaked to third parties as the error correction is done on the public channel. Privacy amplification methods are employed to mitigate this security problem. Privacy amplification is generally achieved by using a set of universal hashing functions to generate a shorter but randomized new key.

### B. The optimization of publicly shared information through the classical channel

Consider two communicating parties that exchange information about photon detection times and measurement bases. An eavesdropper can access measurement results by using them and it is quantified by mutual information between them [11]:

$$H(T) = -\int \bar{d}(t) \log_2[\bar{d}(t)]dt \quad (5)$$

$$H(X) = -\sum_x p^0(x) \log_2[p^0(x)] \quad (6)$$

$$H(X,T) = -\sum_x \int p(x,t) \log_2[p(x,t)]dt \quad (7)$$

$$= -\sum_x \int p^0(x) d_x(t) \log_2[p^0(x) d_x(t)]dt \quad (8)$$

where $x$ is a label that stands for a base in a basis set which represents a measurement basis for one communicating party. All the bases that span the Hilbert space of this basis set should be taken into account, e.g. $x = \{0,1\}$ for rectilinear basis set $(45^0, 135^0) \equiv (0,1)$. $p^0(x)$ is a priori probability of appearing of $x$ in the protocol. $d_x(t)$ is the timing histogram of a detector, namely, the photon counting function of a detector.

In a detection module, there are optical paths at the amount of the number of values of $x$ can take relating to a measurement basis. An optical path starts from the point that a photon enters into the detection module and ends at the point of its detection. Two important parameters that determine the amount of mutual information are $\Delta t_0$ and time bin size. $\Delta t_0$ is the difference between the optical path length + and the centroid location of photon counting function of detectors.

In Fig. 6, there are two detectors, one is shifted at the amounts of $\Delta t_0$ according to the other. These two detectors have the same timing histogram profile other than the centroid locations.

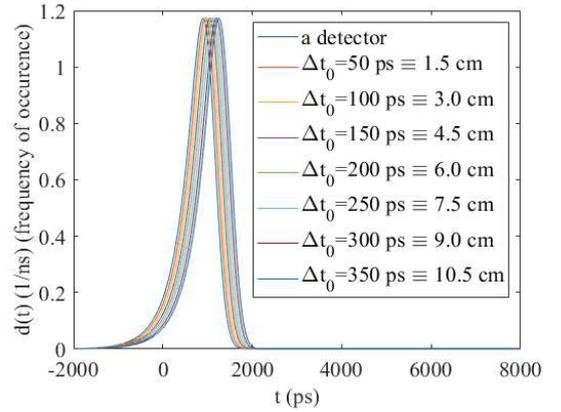

Fig. 6  Detector histograms with different timing jitter profiles

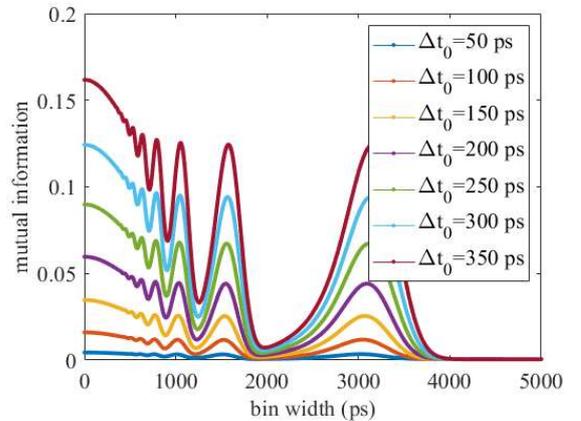

Fig. 7  Change in mutual information with respect to bin width assuming the detection starts at 0 ps.

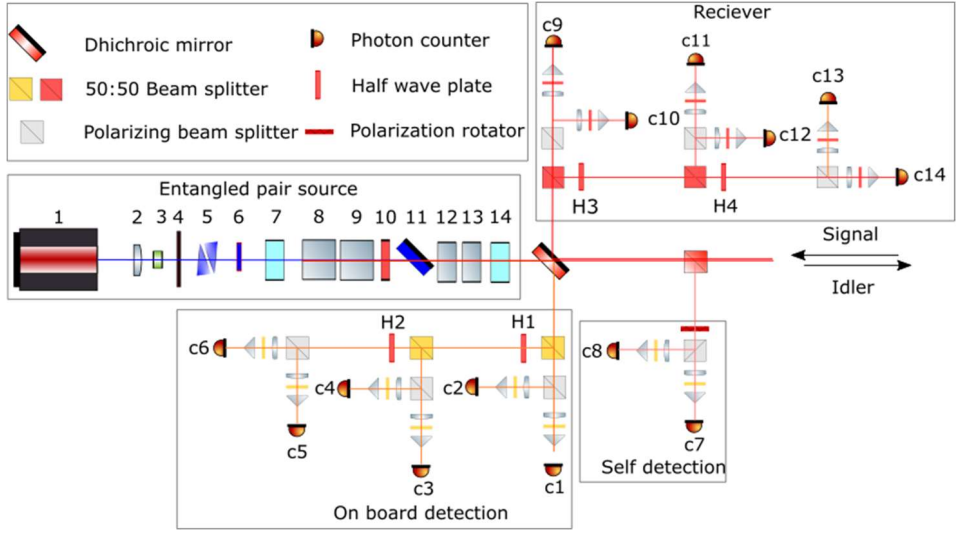

Fig. 8 Experimental arrangement for entanglement based two-way quantum key distribution network. Components inline are 1. Laser, 2. Aspheric lens, 3. Fluorescence filter, 4.Iris, 5. Glan Taylor polarizer, 6. Half-wave plate, 7. Pre-compensator (temporal), 8. BBO1, 9. BBO2, 10. Longpass filter, 11. Dichroic mirror, 12& 13. Post-compensators (spatial), 14. Post-compensator (temporal). A laser source of wavelength 405 nm, after spectral cleaning (2 through 4) and polarized at 45^∘ (5 & 6), interacts with two orthogonally aligned BBO crystals, where horizontally and vertically polarized signal and idler photon are created in BBO1 and BBO2, respectively. The blue pump cut-off from the system by a dichroic mirror. A temporal pre-compensator and post-compensator clean up the temporal information. Two BBO post compensators (12 & 13) are placed to erase the spatial uncorrelation that can reveal which-crystal information. Photon counters and half-wave plates are indicated with letters 'C' and 'H' respectively.

In Fig. 7, change in the mutual information is shown as a function of time bin size for different $\Delta t_0$s. Time bin size parameter is shown as bin widths in the graph. When bin width is scanned along the main distribution of timing histogram, it is seen that peaks occur at some points. In the occurrance of peaks, the starting time of binning in the timing histogram, in other words the position of the main distribution in the whole time interval, is important. Since an absolute decay of mutual information is not seen, the bin width value which will be used in the QKD protocol should be chosen carefully.

The binned versions of timing histogram corresponding to the minimum and maximum values are seen in Fig. 9 and 10 respectively. The goodness and badness of bin widths can be seen in the shape of the binned versions.

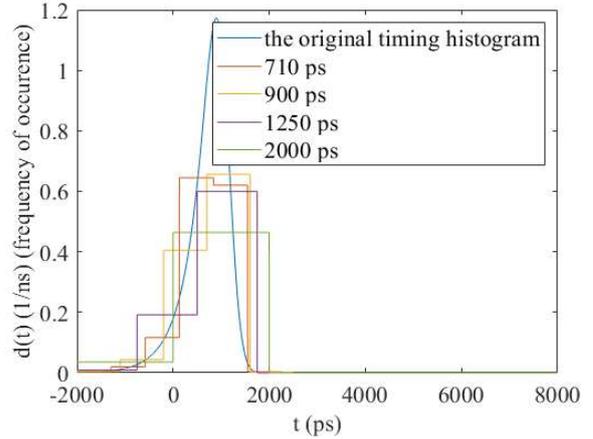

Fig. 10 Detector histograms with minimum mutual information

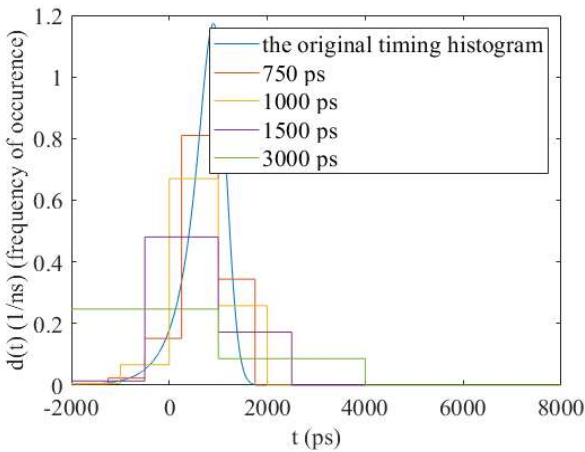

Fig. 9 Detector histograms with maximum mutual information

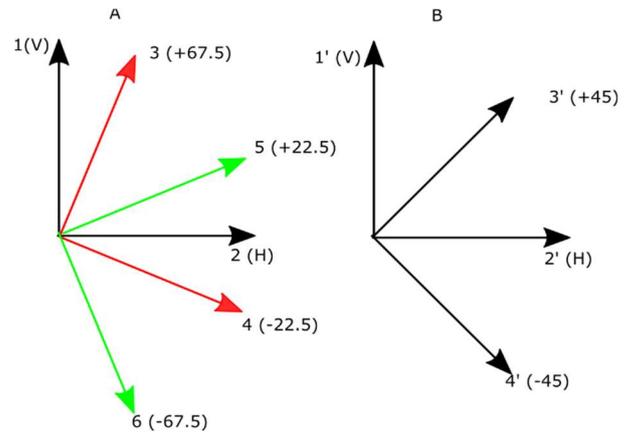

Fig. 11 Measurement bases at Alice and Bob's sides

## C. Quantum Payload Layout

The co-polarized entangled photon pairs, called signal and idler, produced in an SPDC process spectrally separated by a dichroic mirror as shown in Figure.4. Our QKD scheme consists of two entangled photons sources at each party (only one party is shown in Fig. 4 for simplicity, the other party consists of an identical arrangement). The QKD system involves three distinct basis settings $A_0 = \pm 22.5°$, $A_1 = \pm 67.5°$, $A_k = 0°, 90°$ on one side and three basis settings $B_0 = 0°, 90°$, $B_1 = \pm 45°$, $B_k = 0°, 90°$ on the other side for achieving polarization measurement on photon pairs in a singlet state $|\psi\rangle = \frac{1}{\sqrt{2}}(|HH\rangle + |VV\rangle)$. Basis settings $A_k$ and $B_k$ are parallel, and therefore corresponds to horizontal - horizontal polarization detection leads to perfectly correlated results; used for sharing the row key. Other bases are used for performing a violation of CHSH inequality S, with $S = E(A_0, B_0) + E(A_0, B_1) + E(A_1, B_1) - E(A_1, B_0)$.

The value of E is calculated by

$$E(A_0, B_0) = \frac{N_{3,1'} + N_{4,2'} - N_{3,2'} - N_{4,1'}}{N_{3,1'} + N_{4,2'} + N_{3,2'} + N_{4,1'}} \quad (9)$$

*D. Photon detection optimization*

For the On-board single photon detection and counting we considered two main challenges in design. The first challenge is need for high computing power on-board satellite for time-correlation of photons. Our solution for this challenge is based on reconfigurable and fault tolerant use of COTS (commercial off the shelf) FPGAs.

Application of non-radiation hardened FPGAs have been tested in recent years and approved on many CubeSat projects and products [12]. Using techniques like TMR fault tolerance inside the FPGA [9, 10] enables us to use COTS FPGAs in aerospace applications. This design can support possibility of generating hyper-entangled states, where multiple quantum states can be encoded into a single photon. To be able to use both polarization and time-bin degree of freedom an on-board reconfigurable FPGA will be used in design. The on-the-fly reconfiguration ability of the FPGA provides a considerable advantage in power saving and reliability. It means that for the times that we need single-photon timestamp correlation, the FPGA is configured as a Time-Correlated Single Photon Counter (TCSPC) and in the free time slots it will be configured to perform other tasks like soft processor, software radio modem, etc. By using the mentioned advanced techniques, we will be enabled to use COTS FPGAs like Zynq series in our design for both On-board soft computing purposes and hard real-time tasks like Time-Correlated Single Photon Counting. This will enable us to implement intersatellite Quantum links in CubeSat with low power consumption and low cost.

The second challenge in single photon detection is based on rapid temperature changes of the satellite due to exposure to the sun which causes considerable changes in the photon detection efficiency of Avalanche Photo Diodes (APD) that are used for single photon detection. Our solution to this challenge is based on a feedback loop which detects the amplitudes of the analog pulses coming from APD and tunes the High-voltage bias of the APD according to pulse amplitudes. Using this technique, we will stabilize the photon detection efficiency of the on-board single photon detector.

## IV. CONCLUSION

The QKD reached the threshold to become a global network. CubeSat technology seems as the best candidate for achieving this goal. However, the SWaP restrictions impose constraints on other parameters. A careful design is required to overcome the challenges on optical link, data budget and other parameters. Another set of challenges come from the QKD protocol steps. A poor choice of time bin size, geometry of quantum payload or type of detectors may result in possible side channel attacks. However, it is possible to exploit CubeSats for the purpose of a global QKD network with careful design and optimization of parameters.

## ACKNOWLEDGMENT

This work is supported by TUBITAK, project no. 118E991. The authors would like to than Reza Bayat for editing the manuscript.

## REFERENCES

[1] R. Bedington et al., "Nanosatellite experiments to enable future space-based QKD missions," EPJ Quantum Technology, vol. 3, no. 12, Oct., 2016, http://dx.doi.org/10.1140/epjqt/s40507-016-0051-7.

[2] A. Villar et al., "Entanglement demonstration on board a nano-satellite", Optica, vol. 7, no. 7, pp. 734-737, July, 2020, https://doi.org/10.1364/OPTICA.387306

[3] X. Guo, C. Zou, C. Schuck, H. Jung, R. Cheng, and H.X. Tang, (2016) Light: Sci. Appl. 6, e16249 (2017).

[4] Pavel Kovar, "Experiences with the GPS in unstabilized CubeSat," International Journal of Aerospace Engineering (2020).

[5] Yang Yang, et al., "Orbit determination using combined GPS+ Beidou observations for low earth CubeSats: software validation in ground testbed," China Satellite Navigation Conference (CSNC) 2015 Proceedings: vol. III. Springer, Berlin, Heidelberg, 2015.

[6] Erin Kahr, et al., "GPS tracking on a nanosatellite the CanX-2 flight experience," 8th International ESA Conference on Guidance, Navigation & Control Systems, Karlovy Vary, Czech Republic. 2011.\

[7] Stefan Gundacker, and Arjan Heering, "The silicon photomultiplier: fundamentals and applications of a modern solid-state photon detector," Physics in Medicine & Biology, vol. 65 no. 17 (2020): 17TR01.

[8] M. Mehic, et. al, "Error reconciliation in quantum key distribution protocols," in Reversible Computation: Extending Horizons of Computing, I. Ulidowski, I. Lanese, U. P. Schultz, C. Ferreira, Eds., Cham, Switzerland: SpringerOpen, 2020, ch. 6.

[9] T.H. Nguyen, et al., "Reconfiguration control networks for FPGA-based TMR systems with modular error recovery," Microprocessors and Microsystems 60 (2018), pp. 86-95.

[10] Fernanda Kastensmidt, and Paolo Rech, "Radiation effects and fault tolerance techniques for FPGAs and GPUs," FPGAs and Parallel Architectures for Aerospace Applications, Springer, Cham, 2016, pp. 3-17.

[11] A. Lamas-Linares and C. Kurtsiefer, "Breaking a quantum key distribution system through a timing side channel," Opt. Express 15, 9388-9393 (2007).

[12] Tom Neubert, et al., "System-on-module-based long-life electronics for remote sensing imaging with CubeSats in low-earth-orbits," Journal of Applied Remote Sensing, vol.13, no. 3 (2019): 032507